\documentstyle[12pt,amssymb,latexsym,amsmath]{article}
\begin{document}

\title{\bf On  integration of some classes of
$(n+1)$ dimensional
 nonlinear Partial Differential Equations 
}

\author{ Alexandre I. Zenchuk\\
Center of Nonlinear Studies of\\ L.D.Landau Institute 
for Theoretical Physics  \\
(International Institute of Nonlinear Science)\\
Kosygina 2, Moscow, Russia 119334\\
E-mail: zenchuk@itp.ac.ru\\}
\maketitle

\begin{abstract}
The paper  represents the method for construction of the families of
particular solutions to some new classes of $(n+1)$ dimensional 
nonlinear Partial Differential
Equations (PDE). Method
is based on the specific link between algebraic matrix equations
and PDE.
Admittable solutions depend on arbitrary functions of $n$ variables.
\end{abstract}
 
 \section{Introduction}
 Many different methods have
been developed for analytical investigation of nonlinear PDE
during last decades. 
Especially attractive are methods for study of so-called completely
integrable systems.  The
particular interest to these equations is enhanced due to  their 
wide range of application in physics. 
 We emphasize  different dressing
methods, which are based on  fundamental properties of linear
operators,
either differential or integral: 
Zakharov-Shabat dressing method \cite{ZSh1,ZSh2}, $\bar{\partial}$-problem
\cite{ZM,BM,K,Z7}, Sato theory
\cite{OSTT,Z4}.
 
We suggest the method for construction of the families of
particular solutions to some new classes of $(n+1)$ dimensional
nonlinear PDE, $n\ge 2$. It is
 based on general properties of linear {\it algebraic} matrix
 equations. Essentially we 
  develop some ideas represented in the ref.\cite{OSTT}  and
recently in the ref. \cite{Z4,Z8}. In general, for $(n+1)$
dimensional PDE
this method supplies solutions,
depending on the set of arbitrary functions of $n$ variables.
The represented method  works also for classical $(2+1)$-dimensional PDE
integrable by the Inverse Scattering Technique (IST). 
In this case our algorithm is similar to the algorithm
represented in refs.\cite{OSTT,GM1}. 

The structure of the paper is following. First, we discuss
general algorithm relating linear algebraic equation with
nonlinear PDE. Then we show that these PDE are compatibility
condition for some overdetermined linear system of equations 
having different structure in comparison with linear system
associated with completely integrable nonlinear PDE. 
We give an example of (2+1)-dimensional 
system which can not be described in frames of classical dressing
methods.

\section{General algorithm}

As mentioned above,
our  algorithm is based on the
fundamental properties of linear matrix algebraic equation
\begin{eqnarray}\label{MATR}
\Psi U = \Phi,
\end{eqnarray}
where $\Psi=\{\psi_{ij}\}$ is $N\times N$ nondegenerate matrix,
$U$ and  $\Phi$ are $N\times M$ matrices $M<N$.
Namely, the solution of this equation is unique,
$U=\Psi^{-1}\Phi$, and consequently
the homogeneous equation with matrix $\Psi$ has only the trivial
solution.
Thus, if we find
 transformation
$T$ which maps the nonhomogeneous equation (\ref{MATR}) into the
homogeneous equation
$\Psi \tilde U(U) =0
$,
then
$\tilde U(U)=0$.

Let us show  that such transformations can be performed  by
means of
differential operators having special structure.
For this purpose  let us introduce two types of additional parameters 
$x=(x_1,\dots, x_n)$ ($n={\mbox{dim}}(x)$) and  $t=(t_1,t_2,\dots)$
with the following systems:
\begin{eqnarray}\label{x}
\Psi_{x_i} = \Psi B_i + \Phi C_i,\;\;i=1,\dots,n
\end{eqnarray}$
(B_i$ and $C_i$ are  constant $N\times N$ and $M\times N$ matrices
respectively) and
\begin{eqnarray}\label{t}
{\cal{ M}}_i\Psi =0,\;\;\; {\cal{M}}_i\Phi = 0,\;\;
{\cal{M}}_i=\partial_{t_i} + L_i,\;\;
\end{eqnarray}
where $L_i$ are arbitrary linear
differential operators having derivatives with respect to 
variables $x_j$ and
constant scalar coefficients,
so that the system (\ref{x}) is compatible with  the system
(\ref{t}). For the sake of simplicity in this paper we use only 
one parameter $t$, omit subscripts in the eq.(\ref{t}) and use
$n$-dimensional Laplacian for $L$: 
\begin{eqnarray}\label{Mi}
 {\cal{M}}=\partial_{t}+\sum_{k=1}^n\alpha_k\partial_{x_k}^2
\end{eqnarray}
Hereafter indexes $i$, $j$ and $k$ run values from $1$ to $n$ 
unless otherwise specified. 

Let us study compatibility conditions for 
the system (\ref{x}) itself, which 
 has the following form:
\begin{eqnarray}\label{comp}
(\Psi B_j + \Phi C_j)) B_i + \Phi_{x_j} C_i =
(\Psi B_i + \Phi C_i)) B_j + \Phi_{x_i} C_j.
\end{eqnarray}
Require that matrices $B_i$
and $C_i$  satisfy two conditions:
\begin{eqnarray}\label{matrcomp1}\label{matrcomp2}
C_j B_i-C_i B_j=0,\;\; B_jB_i-B_i B_j=0,\;\;i\neq j,
\end{eqnarray}
and matrices $C_j$ have the following structure:
 $C_j=\left[\begin{array}{c|c}
 P_{j}&0_{M,N-R}
\end{array}\right]$,
$R\le M
$, 
 where
  $P_{j}$ are $M\times R$ matrices with rang $R$ and  $0_{A,B}$
  means
  $A\times B$ zero matrix.
 Then equation (\ref{comp}) is reduced to the next one:
 \begin{eqnarray}\label{Phi}
 \Phi_{x_i} P_{j}- \Phi_{x_j} P_{i}=0,
 \end{eqnarray}
 which results  in the first  
nonlinear equation for
$U$ owing to the eq.(\ref{MATR}):
 \begin{eqnarray}\label{comp1}
(B_j+UC_j) UP_{i} + U_{x_j} P_{i} =
(B_i+UC_i) UP_{j} + U_{x_i} P_{j}.
\end{eqnarray}

Let us show  that   another nonlinear matrix equation can be
derived using operator ${\cal{M}}$.
  For this purpose we
apply operator ${\cal{M}}$  to both sides of the eq.
(\ref{MATR}) and use eqs. (\ref{x}) and  (\ref{t}):
\begin{eqnarray}\label{MPsi1} 0={\cal{M}}\Phi =  ({\cal{M}} \Psi) U+ \Psi
{\cal{M}} U + 2 \sum_{k=1}^n \alpha_k\Psi_{x_k} U_{x_k}
=  \Psi \left(
{\cal{M}} U + 2 \sum_{k=1}^n \alpha_i (B_k+ U C_k) U_{x_k} \right) =0 
\end{eqnarray} 
Since ${\mbox{det}}(\Psi)\neq 0$ one  has the second 
nonlinear equation for the  matrix $U$: 
\begin{eqnarray}\label{flow2}
  U_{t} + \sum_{k=1}^n \alpha_k \left(U_{x_kx_k} +2 (B_k+ U
C_k) U_{x_k} \right)  =0.
\end{eqnarray}
Having eqs.(\ref{matrcomp1},\ref{comp1},\ref{flow2}) one can derive the
complete system of equations for  elements of the 
matrix $V$ composed of first $R$ rows of the matrix $U
$. 
First equation exists  if $n>2$. In this case we
can write the   matrix equation without operators $B_i$
 using any three equations (\ref{comp1})  with pairs of
indexes $(i,j)$, $(j,k)$ and $(k,i)$ and relations
(\ref{matrcomp1}) 
($U_i=V P_i$): 
\begin{eqnarray}\label{nonlin1}
\sum_{perm} P_i \left({U_j}_{x_k} - {U_k}_{x_j} +
[U_k,U_j]\right)=0,
\end{eqnarray}
where sum is over cycle permutation of indexes $i$, $j$ and
$k$.
In general,
this equation is not complete system for all elements of the
matrix $V$. To complete it we derive additional matrix
equation, which can be done for any $n$.  
 Eliminate operators $B_i
 $ from the eq.
(\ref{comp1}) ($i=1$, $j=2$)  using eqs. 
(\ref{matrcomp1}) and (\ref{flow2}):
\begin{eqnarray}\label{nored}
P_i{U_j}_t - P_j {U_i}_t+
2 \sum_{k=1,k\neq i}^n\alpha_k(P_i U_k-P_kU_i){U_j}_{x_k}-
2 \sum_{k=1,k\neq j}^n\alpha_k(P_j U_k-P_kU_j){U_i}_{x_k}+\\\nonumber
\sum_{k=1}^n\alpha_k \left(P_i{U_j}_{x_kx_k}-P_j {U_i}_{x_kx_k}+
2 P_k({U_i}_{x_j} - {U_j}_{x_i})_{x_k}-2 P_k ({U_i}_{x_k} U_j-
{U_j}_{x_k} U_i)
\right)=0
\end{eqnarray}
In particular, if
$\alpha_j=0$, $j>1$, $\alpha_1=1$, then 
eq.(\ref{nored}) reduces in
\begin{eqnarray}\label{red1}
P_1 \left({U_2}_{t} + 2 {U_1}_{x_1x_2} - {U_2}_{x_1x_1}+
2 (U_2U_1)_{x_1} - 2 {U_1}_{x_1} U_2
\right)-\\\nonumber
P_2 \left({U_1}_{t} +  {U_1}_{x_1x_1} + 2 U_1 {U_1}_{x_1}
\right)=0.
\end{eqnarray} 
Thus equations (\ref{nonlin1},\ref{nored},\ref{red1})
do not depend on both parameter $N$ (which characterizes dimensions of
the matrices in the eq.(\ref{MATR})) and matrices $B_i$, i.e. 
$N$ is arbitrary positive integer, $B_i$ are arbitrary $N\times N$ matrices 
fitting relations (\ref{matrcomp1}).   

We point on two trivial 
reductions of the eq.(\ref{red1}).
 
1. Let 
 matrix $A$ exist such that $A P_1=0$  and
$AP_2=I_R$  ($I_R$ is $R\times R$ identity matrix).
Multiplying eq.(\ref{red1}) by $A$ from the right
one receives matrix Burgers equation for $U_1$. 

2. If $R=M=2$ and 
 \begin{eqnarray}\label{P}
 P_1=I_2, \;\;
P_2=\left[\begin{array}{cc}
0&1\cr1&0\end{array}\right],\;\; V=
\left[\begin{array}{cc}
u_1&v_1\cr u_2&v_2\end{array}\right]
,
 \end{eqnarray}
 then eq.(\ref{red1}) is reduced to the next system:
\begin{eqnarray}\label{DS}
 r_{t} - r_{x_1x_2} -2 r w_{x_1x_2} =0,\;\;\;
 q_{t} + q_{x_1x_2} +2 q w_{x_1x_2} =0 ,\;\;\;
 w_{x_1x_1} - w_{x_2x_2} = qr ,
 \end{eqnarray}
 where functions $r$ and $q$ are related with elements of the
 matrix $V$ by the formulae
 \begin{eqnarray}
u_1=\frac{1}{4} (r-q+2 w_{x_1}),\;\;u_2=\frac{1}{4} (r+q+2
w_{x_2}),\;\;\\\nonumber
v_1=\frac{1}{4} (-r-q+2 w_{x_2}),\;\;v_2=\frac{1}{4} (-r+q+2
w_{x_1}).\;\;
\end{eqnarray}
Eq.(\ref{DS}) becomes Devi-Stewartson equation (DS) 
after reduction
$r=\psi$, $q=\bar\psi$, $t_1 = i t$, where $i^2=-1$, bar means complex
conjugated value. 
 
 We will see in
the next section, that the case $ R=M$ does always correspond to the
classical completely
integrable $(2+1)$-dimensional systems.

  Arbitrary functions of variables $x_j$ ($j=1,\dots,n$)
      appear in the solution
  $V$ due to the matrix function $\Phi$, defined by the system (\ref{Phi}).
  The number of arguments in the arbitrary functions as well as
  the number of these functions is defined by particular choice
  of the matrices $P_j$ and dimension $n$ of $x$-space. 
  If $n=2$, then one has at most $N$ arbitrary functions
  of two variables, see {\bf Example}.    
  In general, for    $n$-dimensional $x$-space and $R<M$  we are able to
  represent examples with 
  $N$ functions of $n$ variables. If
  $R=M$ then $\Phi$ may depend at most on  $N\times M$ 
  arbitrary scalar functions of {\it single} variable, which is in accordance
  with \cite{OSTT}.
  Detailed discussion of this
  problem is left beyond the scope of this paper.
 
 
\subsection{On the operator representation of PDE}

We derive the overdetermined linear system of
PDE with compatibility condition in the form of
 eqs.(\ref{nonlin1}) and (\ref{nored}).  
 First,
introduce arbitrary  $R\times N$ matrix function ${\bf R} (\lambda)$
of the additional parameter
$\lambda$.
Multiply eqs.(\ref{comp1}) and (\ref{flow2}) 
by ${\bf R} (\lambda)\exp(\eta) $ from the left and 
introduce function $\hat \Psi = {\bf R}  e^{\eta} U$
\begin{eqnarray}\label{newfunc}\label{eta}
\eta= \sum_{k=1}^n B_k x_k -
 \left(\sum_{k=1}^n  \alpha_k B_k^2\right) t.
\end{eqnarray}
We get after transformations:
 \begin{eqnarray}\label{comp1lin}
  \hat\Psi_{x_j} P_{i}- \hat\Psi_{x_i} P_{j}=
\hat\Psi P_i U_j-\hat\Psi P_j U_i,\\
\label{flow2lin}
  \hat\Psi_{t} + \sum_{k=1}^n \alpha_k \left(\hat\Psi_{x_kx_k} +2
  \hat\Psi P_k V_{x_k} \right)=0.
\end{eqnarray}
If $R=M$, t.e. all $P_j$ are square nondegenerate matrices, then
the system (\ref{comp1lin}), (\ref{flow2lin}) is equivalent to
the classical $M\times M$ overdetermined linear system 
for correspondent 
(2+1)-dimensional integrable system. 
In fact, one can express all derivatives 
of $\hat\Psi$ with
respect to $x_j$, $j>1$ through the derivatives of $\hat\Psi$ 
with respect to 
$x_1$ using equation (\ref{comp1lin}).
Both equations (\ref{comp1lin}) and (\ref{flow2lin}) are $M\times M$
matrix equations for  $M\times M$ matrix function $\hat\Psi$. Thus  
eq. (\ref{comp1lin}) can be taken for the spectral problem while eq.
(\ref{flow2lin}) represents evolution part of the overdetermined
linear system.
For instance, if $M=R=n=2$, $\alpha_1=1$, $\alpha_k=0$, $k>1$,
 and $P_i$ have the form (\ref{P}), 
then the compatibility condition of the linear system 
(\ref{comp1lin},\ref{flow2lin}) is given by the 
eqs.(\ref{DS}).

In the  case $R<M$ situation is  different.  $\hat \Psi$ is $R\times M$ matrix
function, while (\ref{comp1lin}) is $R\times R$ 
matrix equation.
Thus  it
can not be taken for the spectral problem. Also, eq.(\ref{flow2lin})
involves all derivatives which form operator $M$. So, we have
($n+1$) dimensional equations.
Below we represent example of $(2+1)$-dimensional 
system of this type. 

\section{Example}
Let $M=3$, $R=2$, $Q=2$, $\alpha_2=\alpha_3=0$, $\alpha_1=1$,
 $N=3 k + 2$, $k=1,2,\dots$. Everywhere indexes $i$ and $j$ take
 values $1$ and $2$. Let
\begin{eqnarray}\nonumber
P_1=\left[
\begin{array}{cc}
1&0\cr
0&1\cr
0&0
\end{array}
\right],\;\;P_2=\left[
\begin{array}{cc}
0&0\cr
1&0\cr
0&1
\end{array}
\right],\;\;
V=\left[
\begin{array}{ccc}
u_1&v_1&w_1\cr
u_2&v_2&w_2
\end{array}
\right],
\;\;\\\nonumber
B_j=\left[\begin{array}{ccc}
0_{2,2}&b_{j1}&0_{2,N-5}\cr
0_{3 (k-1) ,2}&0_{3(k-1),3}&b_{j2}\cr
0_{3,2}&0_{3,3}&0_{3,N-5}
\end{array}\right],
b_{11}=\left[\begin{array}{ccc}
1&1&1\cr
0&0&0
\end{array}\right],
\;\;\\\nonumber
b_{21}=\left[\begin{array}{cccc}
0&0&0\cr
1&1&1
\end{array}\right],\;\;
b_{j2}={\mbox{diag}}(\underbrace{A_j,A_j,\dots}_{k-1}),\;\;
j=1,2,\;\;\\\nonumber
A_1=\left[\begin{array}{rrr}
1&0&-1\cr
-1&1&0\cr
0&-1&1
\end{array}
\right],\;\;A_2=\left[\begin{array}{rrr}
1&-1&0\cr
0&1&-1\cr
-1&0&1
\end{array}
\right],
\end{eqnarray}
$\Phi=[\phi_1 \;\;\phi_2\;\;\phi_3]$, where
$\phi_k$ are $N$-dimensional columns.
Then equation (\ref{Phi}) 
can be written in the form
${\phi_1}_{x_2}={\phi_2}_{x_1}$, 
${\phi_2}_{x_2}={\phi_3}_{x_1}$, i.e. 
$\phi_1=S_{x_1x_1}$, $\phi_2=S_{x_1x_2}$, 
$\phi_3=S_{x_2x_2}$, where $S$ is arbitrary function of 
variables $x_1$ and $x_2$.
In view of eq.(\ref{t}), we can write for $\Phi$:
$\Phi=\int_{-\infty}^\infty c(k_1,k_2) [1\;\;\;k_2/k_1\;\;\;
 k_2^2/k_1^2] \exp[k_1 x_1 + k_2
x_2 -k_1^2 t] dk_1 dk_2$, where $c(k_1,k_2)$ is arbitrary column
of $N$ elements.
Function $\Psi$ solves system (\ref{x},\ref{t}) and 
can be represented in the form
$
\Psi=(\Psi_0+\tilde \Psi ) e^\eta,
$
where $\eta$ is given by the eq.(\ref{eta}),
  $
\tilde \Psi =\partial_{x_1}^{-1} \Phi C_1 e^{-\eta} 
$ and $\Psi_0$ is arbitrary constant $N\times N$ matrix.
Then eq.(\ref{MATR}) can be solved for the function $U$, its  first $R$ rows 
 define matrix $V$. 
Elements of this matrix
satisfy
the system (\ref{red1}) 
which gets the following form:
\begin{eqnarray}\nonumber 
{u_1}_t +{u_1}_{x_1x_1}- 2
{u_2}_{x_1x_2} + 2 {v_2}_{x_1x_1}+ 2 u_1 {u_1}_{x_1} -
 2 (v_2 {u_1})_{x_1} +&&\\\nonumber 
 2 v_1
{u_2}_{x_1} - 2 (w_2 {u_2})_{x_1} +
 2 (u_2 {v_1})_{x_1} + 4 v_2 { v_2}_{x_1}  &=&0
\\\nonumber 
{v_1}_t-
v_{x_1x_1}+2 {u_1}_{x_1x_2}  + 2 ({w_1} {u_2})_{x_1} + 2 u_1 {v_1}_{x_1}-
2 v_2 {v_1}_{x_1}  &=&0\\\nonumber 
{w_1}_t -
{w_1}_{x_1x_1} + 2 {v_1}_{x_1x_2} - 2 w_1 {u_1}_{x_1}+4 v_1 {v_1}_{x_1} -
2 w_2 {v_1}_{x_1}
+2 (w_1{v_2})_{x_1} &=&0\\\nonumber 
{u_2}_t +{u_2}_{x_1x_1}+ 2 u_2 {u_1}_{x_1} + 2 v_2
{u_2}_{x_1}\;=\;0,\;\;\;
{v_2}_t +{v_2}_{x_1x_1}+ 2 u_2 {v_1}_{x_1} + 2 v_2
{v_2}_{x_1}&=&0\\\nonumber 
{w_2}_t-{w_2}_{x_1x_1} +
2 {u_1}_{x_1x_2}  + 2 {v_2}_{x_1x_2} - 2
{v_1}_{x_1x_1}+ 2 u_2 {w_1}_{x_1} + 2 v_2
{w_2}_{x_1} &=&0\\\nonumber 
\end{eqnarray}

\section{Conclusions}

The represented version of the dressing method 
serves for wide class of $(n+1)$-dimensional PDE. It supplies
solutions depending on arbitrary functions of $n$ variables
provided $R<M$.  In the last case solution depends on the
functions of single variables.
  By construction, equations  have infinite number of
commuting flows and  are
compatibility conditions for some specific  linear overdetermined
systems,  which is  equivalent to the 
classical linear problem if only $R=M$. 

The work 
is supported by RFBR grants 01-01-00929 and 00-15-96007.
Author thanks Prof. S.V.Manakov  and Dr.Marikhin for useful discussions.

\end{document}